\documentclass[prl,twocolumn,nofootinbib,preprintnumbers]{revtex4}

\usepackage{graphicx}
\usepackage{mathrsfs}
\usepackage{amsmath}
\usepackage{hyperref}

\begin{document}

\title{Standard Model Prediction for Cosmological 21cm Circular Polarization}
\author{Lingyuan Ji}
\email{lingyuan.ji@jhu.edu}
\author{Marc Kamionkowski}
\email{kamion@jhu.edu}
\affiliation{Department of Physics and Astronomy, Johns Hopkins University\\ 3400 North Charles Street, Baltimore, MD 21218, United States}
\author{Keisuke Inomata}
\email{inomata@resceu.s.u-tokyo.ac.jp}
\affiliation{Research Center for the Early Universe (RESCEU), Graduate School of Science, The University of Tokyo, Hongo 7-3-1 Bunkyo-ku, Tokyo 113-0033, Japan}

\preprint{RESCEU-6/20}

\begin{abstract}
Before cosmic reionization, hydrogen atoms acquire a spin polarization quadrupole through interaction with the anisotropic 21-cm radiation field. The interaction of this quadrupole with anisotropies in the cosmic microwave background (CMB) radiation field gives a net spin orientation to the hydrogen atoms.
The 21-cm radiation emitted by these spin-oriented hydrogen
atoms is circularly polarized.
Here, we reformulate succinctly the derivation of the expression
for this circular polarization in terms of Cartesian (rather
than spherical) tensors. We then compute the angular power
spectrum of the observed Stokes-$V$ parameter in the standard $\Lambda$CDM cosmological model and show how it depends on redshift, or equivalently, the observed frequency.
\end{abstract}

\maketitle

The redshifted 21-cm line of neutral hydrogen provides the most promising probe of the cosmological ``dark ages,'' the epoch after CMB photons are emitted and before the first stars are formed. While the majority of theoretical work has focussed on intensity fluctuations of the 21-cm radiation \cite{Loeb:2003ya,Furlanetto:2006jb,Lewis:2007kz,Pritchard:2011xb,Morales:2009gs}, there has also been some work on the linear polarization \cite{Babich:2005sb,De:2013wca}.

The {\it circular} polarization of the redshifted 21-cm line was considered in Refs.~\cite{Hirata:2017dku,Mishra:2017lpz}.
Ref.~\cite{Hirata:2017dku} showed that circular polarization arises from an interaction between CMB anisotropies and the atom's spin polarization induced by anisotropies in the 21-cm radiation incident on the atom. 
Ref.~\cite{Mishra:2017lpz} then focussed on the circular polarization from the CMB quadrupole induced by primordial gravitational waves and discussed the prospects to detect an inflationary gravitational-wave background in this way.

In this Letter, we translate the central atomic-physics results of Ref.~\cite{Hirata:2017dku}, which were presented in terms of spherical tensors, in terms of more intuitive Cartesian tensors. We then calculate the angular power spectrum for the 21-cm polarization that arises at second order in the primordial-density-perturbation amplitude in the standard $\Lambda$CDM cosmological model.  We employ aspects of the total-angular-momentum (TAM) formalism \cite{Dai:2012bc,Dai:2012ma} to derive the results in a relatively economical fashion.  We then evaluate the circular-polarization angular power spectrum numerically and determine its dependence on the observed frequency, or equivalently, the redshift of the emitter.
Throughout this letter, we use units in which $c = \hbar = 1$.

Consider the circular polarization $V(\chi,\hat n)$ of the 21-cm radiation that arrives to us from a comoving distance $\chi$ and direction $\hat n$.  The hydrogen atoms at the point $\vec x =\chi\hat n$ are immersed in a 21-cm radiation field that has anisotropies arising from local gas-density inhomogeneities.  This then induces a spin-polarization tensor with a quadrupole aligned with the quadrupole of the 21-cm radiation.  
The atoms are also immersed in a CMB radiation field that also has anisotropies, which are mainly determined by the density fluctuations on the last scattering surface.
Ref.~\cite{Hirata:2017dku} shows that a net spin orientation of the neutral hydrogen arises from the misalignment of the atomic spin-polarization quadrupole and the CMB quadrupole, which leads to spontaneous and stimulated emission of 21cm radiation in direction $-\hat n$ that is circularly polarized.
The spin quadrupole moment of the hydrogen atoms at comoving position $\vec x$ (at the conformal time $\eta=\eta_0-\chi$, where $\eta_0$ is the conformal time today) can be represented as a rank-2 tensor $\gamma_{ab}(\vec x,\eta)$ that is symmetric ($\gamma_{ab}=\gamma_{ba}$) and trace-free ($\gamma_{bb} =0$) --- it is given explicitly in Eq.~(3) of Ref.~\cite{Hirata:2017dku}.  Likewise, the CMB temperature quadrupole at that point is $t_{ab}(\vec x,\eta)\equiv \int d^2 \hat u\, (3 \hat u_a \hat u_b - \delta_{ab}) \Theta(\vec x,\hat u,\eta)$, where  $\Theta(\vec x, \hat u, \eta) \equiv T(\vec x,\hat u,\eta)/T_\gamma(\eta)-1$ with $T(\vec x, \hat u, \eta)$ the CMB temperature at $(\vec x,\eta)$ arriving from direction $\hat u$ and $T_\gamma(\eta)$ the mean CMB temperature at $\eta$.

The circular polarization is parity-odd and is thus a pseudo-scalar. We therefore infer that the circular polarization must be
\begin{equation}\label{eqn:v-n}
	V(\chi,\hat n)= C(\eta_0-\chi) \epsilon_{abc} n_a \gamma_{bd}(\chi\hat n,\eta_0-\chi)t_{cd}(\chi\hat n,\eta_0-\chi),
\end{equation}
as this is the only pseudo-scalar that can be constructed from $n_a$, $\gamma_{ab}$, and $t_{ab}$, and the Levi-Civita symbol $\epsilon_{abc}$. The coefficient $C$ can be determined by comparing Eq.~(\ref{eqn:v-n}),  Eqs.~(4) and (46) in Ref.~\cite{Hirata:2017dku}, and the translation [e.g., Eq.~(3) in that paper] between spherical and Cartesian tensors. The result is
\begin{equation}
	C(\eta) \equiv \sqrt{\frac{2}{3\pi}}\frac{(1+z) T_s(\eta) K_\mathrm{mag}\tau(\eta)}{A[1+0.75\tilde x_\alpha(\eta)]},
\end{equation}
where $K_{\rm mag}=1.65\times 10^{-12}\,\mathrm{s}^{-1}$; $A=2.86\times 10^{-15}\,\mathrm{s}^{-1}$ is the 
Einstein coefficient of the hyperfine transition; $z$ the redshift at conformal time $\eta$; $T_s(\eta)$ the spin temperature at that time; and $\tilde x_\alpha(\eta)$ the coefficient describing the rate of de-alignment of polarized hydrogen atoms. Here, $\tau(\eta)$ is the optical depth in the 21-cm line.

The next step is to determine the connection between the spin-polarization tensor $\gamma_{ab}(\vec x,\eta)$ and the linear-theory fractional density perturbation $\delta(\vec x,\eta)$ at that time.  This tensor can again be written in terms of spherical tensors, and the spherical-tensor components induced by one Fourier mode $\tilde \delta(\vec k,\eta)$, of wavevector $\vec k$, of the density field are [Eq.~(4) in Ref.~\cite{Hirata:2017dku}],
\begin{equation}\label{eqn:p-2m}
	\tilde {\mathscr{P}}_{2m}(\vec k,\eta) = \sqrt{\frac{4\pi}{5}} D(\eta) \tilde \delta(\vec k,\eta) Y_{2m}(\hat k),
\end{equation}
where $Y_{lm}$ are the spherical harmonics.
Here,
\begin{equation}
	D(\eta) \equiv
        \frac{1}{20\sqrt{2}} \frac{T_\star}{T_\gamma(\eta)}
        \left[1 - \frac{T_\gamma(\eta)}{T_s(\eta)} \right]
        \frac{f\tau(\eta)}{1+\tilde  x_\alpha(\eta) + \tilde x_c(\eta)} ,
\end{equation}
is a $\vec k$-independent quantity, where $T_*=68$~mK is the hyperfine splitting in temperature unit, $f$ is the growth rate of structure (which is unity during the matter domination), and $\tilde x_c(\eta)$ describes the rate of collisions with other hydrogen atoms. 
In terms of Cartesian tensors, the relation must take the form,
\begin{equation}
     \gamma_{ab}(\vec x,\eta) = F(\nabla^2) \left(\nabla_a \nabla_b -
     \frac13 \delta_{ab} \nabla^2 \right) \delta(\vec x,\eta),
\end{equation}
(where $\nabla_a \equiv \partial / \partial x_a$) given that any symmetric trace-free rank-2 tensor constructed from the scalar $\delta(\vec x)$ must be proportional to $[\nabla_a \nabla_b - (\delta_{ab}/3)\nabla^2]\delta(\vec x)$.  In Fourier space, this relation becomes
\begin{equation}\label{eqn:fourierspace} 
    \tilde \gamma_{ab}(\vec k, \eta) = -F(-k^2) \left( k_a k_b - \frac{k^2}{3} \delta_{ab} \right) \tilde \delta(\vec k, \eta).
\end{equation}
The function $F(x)$ can be determined, for example, by taking $\vec k = k \hat z$ (which makes $\gamma_{ab}$ diagonal) and comparing Eq.~(\ref{eqn:fourierspace}) with Eq.~(\ref{eqn:p-2m}). Doing so, we find $F(-k^2)=-D/(\sqrt{2} k^2)$, or
\begin{equation}\label{eqn:gamma-delta}
     \tilde \gamma_{ab}(\vec k,\eta) = \frac{D(\eta)}{\sqrt{2}} \left(
     \hat k_a \hat k_b - \frac{\delta_{ab}}{3} \right) \tilde \delta(\vec k,\eta).
\end{equation}

We now review the relation between the CMB-anisotropy tensor $t_{ab}(\vec x,\eta)$ and $\delta(\vec x,\eta)$.  Since $t_{ab}(\vec x,\eta)$ is the quadrupole moment of the CMB anisotropy observed at $(\vec x, \eta)$ during the matter-dominated era, we can obtain it from the Sachs-Wolfe effect. Thus,
\begin{align}
     t_{ab}(\vec x,\eta) =&  \int d^2\hat u\, (3 \hat u_a \hat u_b
     - \delta_{ab}) \Theta(\vec x,\hat u, \eta) \nonumber \\
     =& -\int d^2 \hat u\, \left(\hat u_a \hat u_b - \frac{\delta_{ab}}{3} \right) \Phi[\vec x + \hat u(\eta - \eta_\text{ls}), \eta_\text{ls}],
\end{align}
where $\Phi(\vec x,\eta_{\rm ls})$ is the Newtonian-gauge gravitational potential \cite{Dodelson:2003ft} at the conformal time $\eta_{\rm ls}$ of the CMB surface of last scatter. Using the shift formula, this relation can be written in Fourier space as 
\begin{equation}
	\tilde t_{ab}(\vec k, \eta) = -\left[ \int d^2 \hat u\, \left(\hat u_a \hat u_b - \frac{\delta_{ab}}{3} \right) e^{i \vec k \cdot \hat u (\eta - \eta_\text{ls})}\right] \tilde\Phi(\vec k, \eta_\text{ls}).
\end{equation}
The integral over $\hat u$ can be evaluated by using the plane-wave expansion and taking $\vec k = k\hat z$. The gravitational-potential perturbation can be related to the density perturbation through the (Fourier-space) Poisson equation, $ (k/a)^2 \tilde \Phi(\vec k,\eta) = 4 \pi G \bar \rho \tilde \delta(\vec k,\eta)$, with $a$ the scale factor and $\bar \rho = 3H_0^2 \Omega_m/(8\pi G a^3)$ the mean density in the matter dominated era. Finally, we arrive at
\begin{equation}\label{eqn:t-delta}
     \tilde t_{ab}(\vec k,\eta) = \frac{6\pi H_0^2 \Omega_m}{a(\eta_\text{ls}) k^2}j_2 [k(\eta-\eta_{\rm ls})] \left( \hat k_a \hat k_b -\frac{\delta_{ab}}{3} \right) \tilde \delta(\vec k, \eta_\text{ls}),
\end{equation}
where $j_J(x)$ are the spherical Bessel functions.

Next, we relate the matter perturbation $\delta(\vec x, \eta)$ at conformal time $\eta = \eta_0 -\chi$ and $\eta_\text{ls}$ to the primordial curvature perturbation $\mathcal R(\vec x)$ generated during inflation. Since both times are within the matter dominated era, the relation in Fourier space takes the form \cite{Dodelson:2003ft}
\begin{equation}\label{eqn:delta-curvature}
	\tilde \delta(\vec k, \eta) = \frac{2k^2}{5H_0^2\Omega_m} T(k) D_+(\eta) \tilde {\mathcal R}(\vec k).
\end{equation}
Here $D_+(\eta)$ is the linear structure growth function and $T(k)$ is the matter transfer function normalized to unity at large scales.

We now expand the primordial curvature perturbation
\begin{equation}\label{eqn:scalar-decomp}
	\mathcal R(\vec x) = \sum_{kJM} \mathcal R_{kJM} \left[4\pi i^J
        \Psi^k_{(JM)}(\vec x) \right],
\end{equation}
in terms of scalar TAM waves $\Psi^k _{(JM)}(\vec x) \equiv j_J(k x)Y_{JM}(\hat x)$. Here $\sum_k$ is a shorthand for $\int k^2 dk / (2\pi)^3$.
We assume that $\mathcal R(\vec x)$ is a statistically homogeneous and isotropic random field in which case,
\begin{equation}
\label{eqn:scalar-spectrum}
	\left\langle (\mathcal R_{kJM})^* \mathcal R_{k'J'M'}
	\right \rangle = \delta_{kk'}\delta_{JJ'} \delta_{MM'} P_{\mathcal R}(k),
\end{equation}
where the angle brackets denote an average over all realizations of the random field. Here $\delta_{kk'}$ is a shorthand for $(2\pi)^3 \delta_D(k-k')/k^2$, and $P_{\mathcal R}(k)$ is the primordial curvature power spectrum.

Likewise, a symmetric trace-free tensor field $h_{ab}(\vec x)$ can be expanded in much the same manner
\begin{equation}
\label{eqn:stf-tensor-decomp}
	h_{ab}(\vec x) = \sum_\alpha \sum_{kJM} h^\alpha _{kJM}
        \left[4\pi i^J \Psi^{k,\alpha}_{(JM)ab}(\vec x)
        \right],
\end{equation}
where the sum on $\alpha$ is over the five types ($\alpha = \text{L}, \text{VE}, \text{VB}, \text{TE}, \text{TB}$) of tensor TAM waves. Given that we are here concerned only with primordial density perturbations, we will require only the longitudinal ($\text{L}$) mode which can be obtained from the scalar TAM wave from
\begin{equation}
     \Psi^{k,\text{L}}_{(JM)ab}(\vec x) = \frac{1}{k^2} \sqrt{\frac32}
     \left( \nabla_a \nabla_b - \frac13 \delta_{ab} \nabla^2 \right)
     \Psi^k_{(JM)}( \vec x).
\end{equation}     
These tensor TAM waves can be written in terms of radial functions $R^{\text{L}\beta}_J(kx)$ and tensor spherical harmonics $Y^{\beta}_{(JM)ab}(\hat x)$ as,
\begin{equation}
	\Psi^{k, \text{L}}_{(JM)ab}(\vec x) = \sum_{\beta} R^{\text{L}\beta}_J
        (kx) Y^{\beta}_{(JM)ab}(\hat x).
\end{equation}
Here, the sum on $\beta$ is over $\beta=\{\text{L}, \text{VE}, \text{TE}\}$, the radial eigenfunctions can be inferred from Eq.~(94) in Ref.~\cite{Dai:2012bc}, and the tensor spherical harmonics are defined in Eq.~(91) in that paper.

Thus, we can write
\begin{equation}\label{eqn:gamma-expansion}
     \gamma_{ab}(\vec x,\eta) = \sum_{kJM} T_\gamma(k,\eta) \mathcal R_{kJM}
     \left[4\pi i^J\Psi^{k,\text L}_{(JM)ab}(\vec x)\right],
\end{equation}
and similarly for $t_{ab}$ in terms of a temperature transfer function $T_t(k,\eta)$. The transfer functions $T_\gamma(k,\eta)$ and $T_t(k,\eta)$ can be determined by combining Eqs.~(\ref{eqn:gamma-delta}) and (\ref{eqn:t-delta}) with Eq.~(\ref{eqn:delta-curvature}), then using Eqs.~(8) and (104) of Ref.~\cite{Dai:2012bc} to convert the Fourier amplitudes to TAM coefficients. Doing so, we find
\begin{align}
T_\gamma(k, \eta) &= - \frac{2k^2 D(\eta)}{5\sqrt{3} H_0^2 \Omega_m} T(k) D_+(\eta), \label{eqn:transfer-gamma} \\
T_t(k, \eta) &= - \frac{12\sqrt{2}\pi}{5\sqrt{3}} j_2[k(\eta -\eta_\text{ls})] T(k) \frac{D_+(\eta_\text{ls})}{a (\eta_\text{ls})}.\label{eqn:transfer-t}
\end{align}

In 21-cm measurements, the light received in a given frequency band corresponds to light emitted over a corresponding range of redshifts or, equivalently, comoving distances.  Here we surmise that the circular polarization is measured in a frequency interval that corresponds to emission from a shell of comoving-distance width $\Delta \chi$ centered at $\chi$.  The observed circular polarization in direction $\hat n$ will then be
\begin{align}\label{eqn:v-n-obs} 
       V_\chi (\hat n) &\equiv \int_{\chi-\Delta\chi/2}^{\chi+\Delta \chi/2}
     \frac{d\chi'}{\Delta \chi}\, V(\chi',\hat n) \nonumber \\
     &\simeq C(\eta_0-\chi) \epsilon_{abc} n_a
     \gamma_{bd}(\chi\hat n,\eta_0-\chi) t_{cd}(\chi\hat n,\eta_0-\chi).
\end{align}
The approximation in the second equality will hold as long as the redshift evolution of $t_{ab}$, $\gamma_{ab}$ and $C$ is relatively slow over the integration interval. Below we will assume this approximation is valid and then limit our attention to the circular polarization on angular scales $\theta \gtrsim \Delta \chi/\chi$ (or multipole moments $l \lesssim \chi/\Delta \chi$).

\begin{widetext}
Inserting Eq.~(\ref{eqn:gamma-expansion}) and the similar one for $t_{ab}$ into Eq.~(\ref{eqn:v-n-obs}), the spherical-harmonic expansion coefficients $V_{\chi\,(lm)} \equiv \int d^2\hat n\, Y_{lm}^*(\hat n) V_\chi(\hat n)$ then evaluate to
\begin{multline}
      V_{\chi\,(lm)} = C(\eta_0 - \chi)\sum_{(kJM)_\gamma}\sum_{(kJM)_t}
      \mathcal R_{(kJM)_\gamma} \mathcal R_{(kJM)_t}
      T_\gamma(k_\gamma, \eta_0 - \chi) T_t(k_t, \eta_0 - \chi) \\
       \times \left[
            \tilde R_{J_\gamma}^\text{VE}(k_\gamma \chi) \tilde R_{J_t}^\text{VE}(k_t \chi)K^{\text{VB},\text{VE}}_{lm(JM)_\gamma(JM)_t}
           +\tilde R_{J_\gamma}^\text{TE}(k_\gamma \chi) \tilde R_{J_t}^\text{TE}(k_t \chi)K^{\text{TB},\text{TE}}_{lm(JM)_\gamma(JM)_t}
              \right],
\end{multline}
where $\tilde R^\beta_J(x)\equiv 4\pi i^J R^{\text{L}\beta}_J(x)$.  Here, $K^{\text{VB},\text{VE}}_{lm(JM)_\gamma(JM)_t}$ and $K^{\text{TB},\text{TE}}_{lm(JM)_\gamma(JM)_t}$ involve Wigner-3$j$ symbols and can be inferred from the integrals in Eqs.~(64) and (66), respectively, of Ref.~\cite{Dai:2012ma}; they are nonzero only for $l+J_\gamma+J_t = \text{odd}$. Using Wick's theorem and Eq.~(\ref{eqn:scalar-spectrum}) for the primordial power spectrum, plus the summation properties of the $K$-factors, the angular circular-polarization power spectrum $C_l^{V_\chi V_\chi} \equiv \langle |V_{\chi\,(lm)}|^2 \rangle$ is found to be
\begin{multline}\label{eqn:angular-power-spectrum}
	 C_l^{V_\chi V_\chi} = |C(\eta_0 - \chi)|^2 \sum_{J_\gamma J_t(\text{odd})}\, \sum_{k_\gamma k_t}\,
	 P_{\mathcal R}(k_\gamma) P_{\mathcal R}(k_t)
     \left\{
        T_\gamma(k_\gamma) T_t(k_t) \Big[T_\gamma(k_\gamma) T_t(k_t) - T_\gamma(k_t) T_t(k_\gamma) \Big]
     \right\}_{\eta=\eta_0 - \chi} \\
     \times \frac{(2J_\gamma+1)(2J_t+1)}{4\pi}
     \left| \tilde R_{J_\gamma}^\text{VE}(k_\gamma \chi) \tilde R_{J_t}^\text{VE}(k_t \chi)
     \begin{pmatrix}
     	l & J_\gamma & J_t\\
     	0 & +1 & -1
     \end{pmatrix}
     - \tilde R_{J_\gamma}^\text{TE}(k_\gamma \chi) \tilde R_{J_t}^\text{TE}(k_t \chi)
     \begin{pmatrix}
     	l & J_\gamma & J_t\\
     	0 & +2 & -2
     \end{pmatrix} \right|^2.
\end{multline}
Here ``$J_\gamma J_t(\text{odd})$'' means the sum of all terms where $J_\gamma + J_t + l$ is an odd number for a given $l$, and the combination of transfer functions in the curly brackets should be evaluated at $\eta=\eta_0-\chi$. The all-sky mean-square signal is $\overline{V_\chi^2} \equiv \int d^2 \hat n \, |V_\chi(\hat n)|^2 / (4\pi)$. By Parseval's theorem, it has the expectation value $\langle\overline{V_\chi^2}\rangle = \sum_l (2l+1)C^{V_\chi V_\chi}_l /(4\pi)$.
\newline
\end{widetext}

\begin{figure}
	\includegraphics[width=0.48\textwidth]{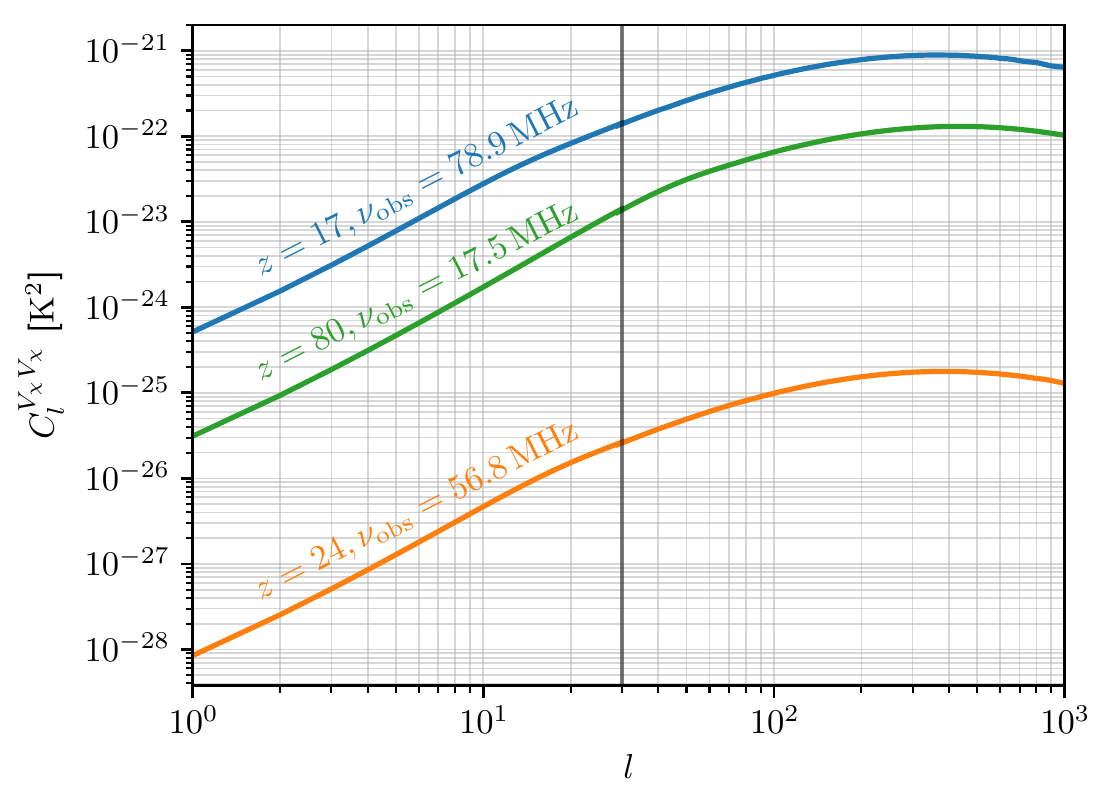}
	\caption{Angular power spectra $C_l^{V_\chi V_\chi}$ of the 21-cm
        circular polarization sourced by scalar perturbation at
        redshifts $z=17$, 24, and 80, corresponding to the redshifted
        frequencies $\nu_\text{obs}=1420/(1+z)\,\text{MHz}=78.9\,\text{MHz}, 56.8\,\text{MHz}, 17.5\,\text{MHz}$.  The signals peak around $l\sim 400$, corresponding to an angular scale $\theta \sim 180^\circ/l \sim 0.5^\circ$.}
	\label{fig:angular-power-spectra}
\end{figure}

Evaluations of the angular power spectrum Eq.~(\ref{eqn:angular-power-spectrum}) for the circular polarizations emitted at three redshifts $z=17, 24, 80$ [redshifted frequencies $\nu_\text{obs}=1420/(1+z)\,\text{MHz}$] are shown in Fig.~\ref{fig:angular-power-spectra}. 
The three redshifts correspond roughly to the times when X-rays from stellar remnants starts to heat the gas ($z \sim 17$), when the Lyman-$\alpha$ photons from the first stars start to heat the hydrogen atoms ($z \sim 24$), and when the spin temperature begins to approach the CMB temperature ($z \sim 80$).  The functions $C(\eta)$ and $D(\eta)$ depend on the details of the ionization history and heating of the IGM and are quite uncertain, particularly at the lower redshifts ($z\lesssim 20$) associated with the epoch of reionization.  To illustrate, we use here the model (designed to be roughly consistent with EDGES \cite{Bowman:2018yin}) detailed in Appendix B of Ref.~\cite{Kovetz:2018zan}.
For low multipoles ($l\leq 30$), Eq.~(\ref{eqn:angular-power-spectrum}) is evaluated exactly using \texttt{WIGXJPF} \cite{Johansson:2015cca} to evaluate the Wigner-3j symbols.  For high multipoles ($l>30$), we use the flat-sky approximation \cite{Hu:2000ee,Inomata:2018vbu}.

\begin{figure}
	\includegraphics[width=0.48\textwidth]{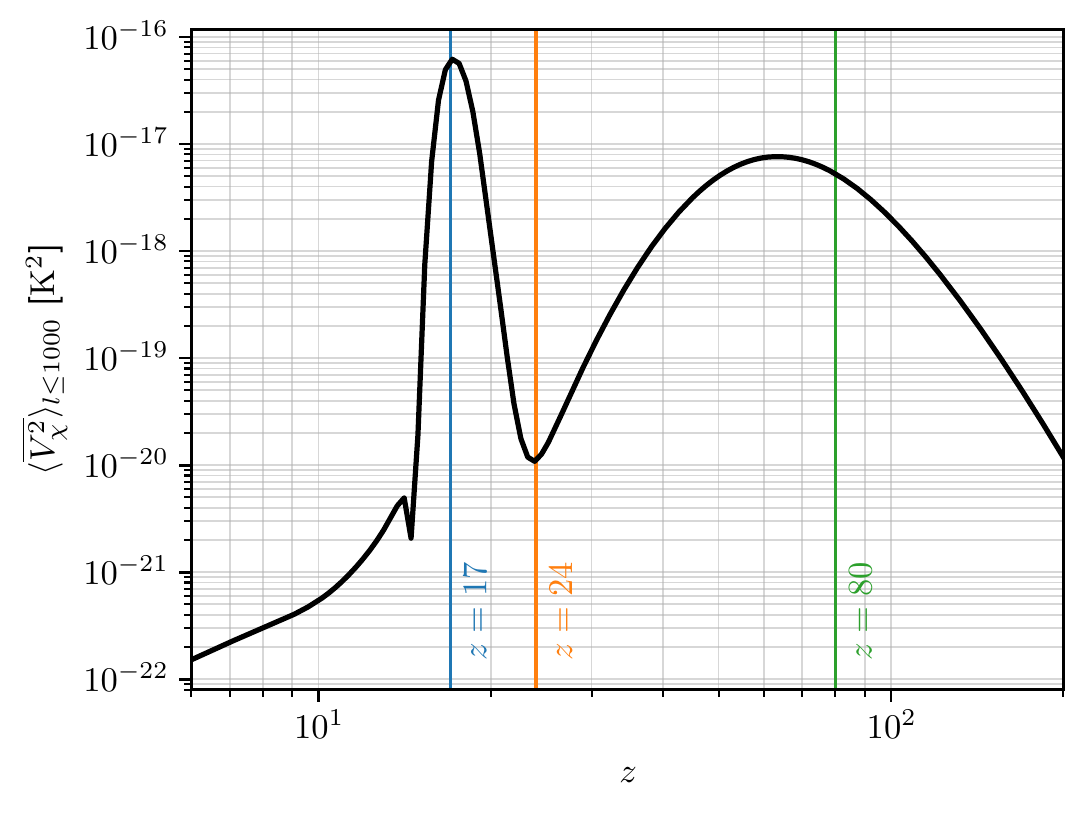}
	\caption{The mean-square signal $\langle \overline{V_\chi^2}\rangle_{l \leq 1000} \equiv \sum_{l=0}^{1000}(2l+1)C_l^{V_\chi V_\chi}/(4\pi)$ as a function of the redshift. The three vertical lines correspond the power spectra plotted in Fig.~\ref{fig:angular-power-spectra}. The strongest signal is sourced around $z=17$, corresponding to a redshifted frequency $\nu_{\rm obs}=78.9\,\text{MHz}$.}
	\label{fig:v-rms-1000}
\end{figure}

The numerical results indicate that the angular power spectra at different redshifts differ primarily in their magnitude, while the angular ($l$) dependence is similar (though not exactly).  The power is spread over a wide range of angular scales but peaks at $l\sim 400$ corresponding to an angular scale $\sim0.5^\circ$.  We indicate the redshift dependence of the signal through the mean-square circular polarization, $\langle \overline{V_\chi^2} \rangle_{l \leq 1000} \equiv \sum_{l=0}^{1000}(2l+1)C_l^{V_\chi V_\chi}/(4\pi)$ shown as a function of redshift in Fig.~\ref{fig:v-rms-1000}.  Again, there are considerable uncertainties in this calculation, primarily at lower redshifts, although the gross features should be reliable.  We see that the signal strength can vary quite rapidly with frequency at frequencies $\nu\simeq80$~MHz corresponding to the beginning of the X-ray heating. The detailed frequency dependence here is, however, uncertain.  There is a more robust prediction for a second, wider, peak at $\nu\simeq 20$~MHz.

To close, we have evaluated the circular polarization of the 21-cm radiation from the dark ages and epoch of reionization that arises at second order in the primordial-density-perturbation amplitude.  We leave a detailed exploration of the detectability of the signal, and strategies for detection, to future work.  Still, the estimates of the signal from the gravitational-wave induced CMB quadrupole~\cite{Mishra:2017lpz} suggest that the signal may be within reach of an ambitious lunar radio base~\cite{Jester:2009dw}; if so, the density-perturbation signal considered here, which is at least $1/r$ ($\gtrsim 14$, $r$: tensor-to-scalar ratio~\cite{Aghanim:2018eyx}) times bigger, should also be within reach.  The techniques described here can also be generalized to models with primordial gravitational waves or vector perturbations.  It will also be interesting in future work to investigate the dependence of the signal on the detailed physics of reionization and to consider cross-correlations of this signal with other observables \cite{Alexander:2019sqb,Alvarez:2005sa,Adshead:2007ij,Tashiro:2008vg}.

\paragraph{Acknowledgements.} We thank E.~D.~Kovetz, J.~L.~Bernal, and K.~Boddy for useful discussions and B.~Wang for providing an ionization-history code. LJ and MK were supported by NSF Grant No.\ 1519353, NASA NNX17AK38G, and the Simons Foundation and KI was supported by JSPS KAKENHI Grant Numbers, 15H02082 and 20H05248.

\end{document}